\documentstyle[aps,multicol]{revtex}
\begin{document}
\title {Nucleation and hysteresis in Ising model: Classical theory versus
computer simulation}
\author {Muktish Acharyya{\footnote 
{E-mail:muktish@thp.uni-koeln.de}} and 
Dietrich Stauffer {\footnote {E-mail:stauffer@thp.uni-koeln.de}}} 
\address {Institute for Theoretical Physics, University of Cologne, 
50923 Cologne, Germany}
\date{\today}
\maketitle
\begin{abstract}
 We have studied the nucleation in 
the nearest neighbour ferromagnetic Ising model, in different ($d$) dimensions,
by extensive Monte Carlo simulation using the heat-bath dynamics. 
The nucleation time ($\tau$) has been studied as a 
function of the magnetic field ($h$) for various system sizes in different
dimensions ($d=2,3,4$). The logarithm of
the nucleation time is found to be proportional to the power
($-(d-1)$) of the magnetic
field ($h$) in $d$ dimensions.
The size dependent crossover from coalescence to nucleation regime
is observed in all dimensions. 
The distribution
of metastable lifetimes are studied in both regions.
The numerical results are compared and found to be consistent
with the classical theoretical
predictions. In two dimensions, we have also studied the dynamical response
to a sinusoidally oscillating magnetic field.
The reversal time is studied as a function of the inverse of the coercive
field. The applicability of the classical nucleation theory to study the
hysteresis and coercivity has been discussed.
\end{abstract}
\vspace {1 cm}
\leftline {\bf Keywords: Ising model, Nucleation, Monte Carlo simulation,
Hysteresis}
\leftline {\bf PACS Numbers: 05.50 +q}
\leftline {{\bf Short title:}{\it Nucleation and hysteresis in Ising model...}}
\bigskip
\leftline {\bf I. Introduction}

The dynamical aspects of Ising models is an active area of modern research.
How does the magnetisation relax towards its equilibrium value, if we start
the dynamics with all spins parallel ? How long is the lifetime of a 
metastable state if a magnetic field is antiparallel to the initial spin
orientation ? Can one answer all these questions in the light of growing
and shrinking droplets ?

What happens if all spins are up in presence of a small opposite magnetic
field and the system is below its critical temperature ($T_c$) ? The 
magnetisation first settles to a metastable state, and then a droplet (of 
overturned spins)
larger than a critical size is formed. As the time passes, this droplet
grows radially and the magnetisation jumps to a negative value. Classical
nucleation theory (CNT) \cite{bd} predicts the logarithm of the
nucleation rate (number of supercritical droplets formed per unit time and
per unit volume) to be asymptotically proportional to $h^{1-d}$ in $d$
dimensions, where $h$ is the magnetic field. This has been
verified in the three-dimensional Ising model by Monte Carlo simulation
\cite{ds}. There are some difficulties in measuring the nucleation rates
by checking how long the magnetisation takes to leave its metastable
value. In the asymptotic limit of field ($h$) going to zero for a finite
lattice size, only one supercritical droplet will be formed and it grows
to cover the whole lattice. This is the proper nucleation regime and the
nucleation rate is the reciprocal of the product of nucleation time and
lattice volume. On the other hand, in the coalescence regime, 
with the lattice size
going to infinity at fixed field ($h$), many such supercritical droplets
will be formed at a time and they grow and coalesce and as a
consequence the magnetisation switches sign. 
This effect, already discussed by Binder and M\"uller-Krumbhaar 
\cite{mul} and mathematically shown by Schonmann \cite{sch}, was
demonstrated by Ray and Wang \cite{rw} for Swendsen-Wang dynamics.

In this paper, we have studied the nucleation in the Ising system by
extensive Monte Carlo simulation (with geometric parallelization) using
heat-bath dynamics. We have also studied the system 
size dependent crossover (from nucleation regime to coalescence regime)
and compared the simulational results with the results of classical
nucleation theory. Here, we mainly reexamine 
(following the earlier works \cite{rikvold,ds})
the validity of classical
nucleation theory by extensive Monte Carlo simulation and obtained better
results by applying modern parallel computational techniques.

Recently, the {\it hysteresis} in the kinetic 
Ising model has gotten much attention in research. Extensive Monte Carlo 
simulation \cite{ac,ma} shows that
the hysteresis loop area behaves as a power law with the frequency in the
low frequency ($\omega \to 0$) limit. 
But for very low frequency, the hysteresis loop area
(approximately equal to four times the coercive field) should vary
logarithmically \cite{dhar}
with the frequency as a consequence of classical nucleation
theory.
To explore the reason of
this mismatch 
(between theoretical prediction and numerical results)
we have studied the dynamical behaviour (coercivity and reversal
time) in the two-dimensional Ising model in presence of a sinusoidally
oscillating magnetic field for sufficiently small frequencies 
(as far as possible now)
and compared
the results with the theoretical predictions (CNT). 
This is the main motivation of our present study.

\bigskip
\leftline {\bf II. Classical nucleation theory}

We review briefly the results of classical nucleation
theory (CNT) far below $T_c$. The equilibrium number 
(per site) $n_s$ of droplets, containing $s$
spins is
$$n_s \sim \exp(-E_s/K_BT) $$
\noindent where $E_s$ is the formation free energy of the droplet of size $s$
and $K_B$ is the Boltzmann constant.
CNT assumes a spherical droplet shape and takes (in $d$-dimension)
$$E_s = -2h s + C_d s^{{d-1} \over d}\sigma(T)$$ 
\noindent where $h$ is the absolute value of the applied magnetic field
and $\sigma$ is the temperature dependent surface tension. The critical
size $s^*$ of a droplet which maximises the free energy, is 
$$ s^* = \left( {{(d-1)C_d \sigma} \over {2dh}} \right)^d$$
\noindent and
$$E_{max} = {{K_d \sigma^d} \over {h^{d-1}}}$$
\noindent where $K_d$ and $C_d$ are
$d$-dependent constants.
The number $n_{s^*}$ of supercritical droplets
$$n_{s^*} \sim \exp(-E_{max}/K_BT) \sim 
\exp(-{{K_d \sigma^d} \over {K_BT h^{d-1}}}),$$
\noindent where the symbol $\sim$ stands for asymptotically proportional
for small fields. The nucleation rate $I$
is proportional to $n_{s^*}$.
In the nucleation regime, where only one supercritical droplet
grows and engulfs the whole system, the nucleation time ($\tau$; the time
required by the system to leave the metastable state) is inversely
proportional to the nucleation rate $I$, 
$$\tau \sim I^{-1} \sim 
\exp({{K_d \sigma^d} \over {K_BT h^{d-1}}}).$$
\noindent In the coalescence regime, many such supercritical droplets
form at about the same time, coalesce and ultimately
form a system-spanning big
droplet. The radius ($\sim s^{1/d}$) of a supercritical droplet
grows linearly with time ($t$), consequently, the number of spins ($s$) in
 a supercritical droplet
will grow as $t^d$. For a steady rate of nucleation, the 
rate of change
of magnetisation is 
$It^d$, for a fixed change ($\Delta m$) in magnetisation during the
nucleation time $\tau$, 
$$\Delta m \sim \int_0^{\tau} I 
t^d dt \sim I \tau^{d+1}.$$ 
\noindent So, in the
coalescence regime, 
$$\tau \sim I^{-1/(d+1)} \sim 
\exp({{K_d \sigma^d} \over {(d+1)K_BT h^{d-1}}}).$$

In an infinitely large system only this coalescence regime is seen.
In this paper, we have performed large-scale simulation 
(using geometric parallelization)
of ferromagnetic
nearest neighbour Ising model
(in 2, 3 and 4 dimensional hypercubic lattices) to verify the
prediction of classical nucleation theory described above, in the
generalization of earlier works \cite{rikvold,ds}.

\bigskip
\leftline {\bf III. Model and simulation scheme}

We have used the standard heat-bath technique to orient the spins (Glauber
kinetics) and started with all spins up in a down field.
Initially the system relaxes
towards a metastable state. It remains for a long time (if the field is
quite small) in the metastable
state and then jumps to the other stable state. 
One such time variation of the magnetisation is depicted in Fig. 1.
We have measured
the lifetime of this metastable state and studied it as a function of
the applied field for various system sizes.

The multispin coding technique has been applied to simulate this updating
process. We have used CRAY-T3E supercomputer having 64 bits per
word. We have stored
16 spins in a computer word (64 bits) and updated a computer word (16 spins)
by a single command. In this sense the updating is parallel and saves
computer memory and time. To improve the efficiency of the updating
process we have used geometric parallelization, where the whole lattice
is distributed over $N_p$ processors. Each processor updates
a rectangular strip ($L\times L/N_p$)
of a square lattice (in $d = 2$). The updated values
of the upper and lower lines of the $n$-th strip 
(by $n$-th proccessor) were passed
via message passing. In periodic boundary conditions, this was done
through a ring-type topology. We have simulated systems of smaller sizes
in SUN workstation.
To produce the strong field regime, coalescence regime
and nucleation regime for a fixed system size, one needs to simulate
a large enough system and allow it to nucleate for a wide range of fields.
In an earlier study in two dimension
\cite{rikvold}, due to smaller system size and nucleation time these
three regimes were less clearly observed.

In the nucleation regime, the true nucleation time is quite large and
fluctuates enormously. 
We have checked the range of metastable values for the 
corresponding range of fields used (in this study)
at a fixed temperature. We define the nucleation time as the time required
by the system to have the magnetisation below a cut-off value 
(chosen below the lowest metastable magnetisation).
This choice is quite 
arbitrary and the results do not depend considerably on the choice of this
cut-off value. 
Due to the huge fluctuations in the nucleation time, to avoid the
waste of computer time we have taken the {\it median}
 nucleation time instead of
taking the (algebraic) mean. One such distribution of nucleation time
is shown in Fig. 2. The fluctuation (width of the distribution) in the
nucleation regime is much larger than that in the coalescence regime.

We have also studied in the same system (in two dimensions only), how the
coercive field 
(value of the field for which the magnetization changes sign)
varies with the reversal time
(the time taken to change the sign of the magnetization)
, when the system is placed
in a sinusoidally ($h(t) = h_0 \cos(\omega t)$) varying magnetic field.
In a small lattice (80$\times$80) we have carried out this simulation and
calculated the coercive field and the reversal time for various frequencies.
Since in the first quarter of the cycle the field is positive and becomes
negative after that, the reversal time will be much higher than the nucleation
time defined above. However, in the static limit ($\omega \to 0$), the
reversal time should be equal to the nucleation time. Since in the
nucleation regime, the nucleation time is very high in comparison with the
inverse of the frequency used, these two times will be same. This will give
some idea about the value of the frequency below which the usual classical
nucleation theory can be safely applied to study the hysteresis and
coercivity \cite{ac,ma,dhar}. A crossover theory from hysteresis 
(large $\omega$) to nucleation (low $\omega$) is given in ref \cite{rikvold}c. 

\bigskip
\leftline {\bf IV. Results}

In our simulation in two dimensions, at
 $J/K_BT = 0.625$, we have obtained the results for
various system sizes ranging from $L = 80$ to $L = 2048$. In Fig. 3 these
results are displayed. Three different regimes, the strong field regime (SFR),
the coalescence regime (CR) and the nucleation regimes (NR) are clearly
identified. The median nucleation time 
$\tau$ (in log scale) is plotted against the inverse
of applied magnetic field. In the coalescence regime the fluctuations are
very small and $\log(\tau)$ behaves linearly with $1/h$ as predicted by the
classical nucleation theory. We have also estimated the slope from the linear
best fit. In the nucleation regime, the fluctuations are quite high. From
the prediction of CNT, the slope ($\log(\tau) \sim 1/h$) in this regime should
be three times higher than that in the coalescence regime. Our data 
show good agreement with this. From the estimated slope, we have calculated
the surface tension $\sigma(T)$ and compared it with the previous estimates.

In two dimensions, at the same temperature, 
also the reversal time (in logscale)
is plotted against the inverse of coercive field in the same plot (Fig. 3).
The range of frequencies we have used is from $6.28 \times 10^{-2}$ to
$1.57 \times 10^{-6}$. The topmost datum in Fig. 3 corresponds to the lowest
frequency.
Here also, the coalescence regimes are shown. The slope 
(for very small frequencies) is same with that
for the static case. The nucleation regime is not very clear, however the
data show a tendency towards the nucleation regime and in the long run, we
believe, it could merge with the static nucleation regime. However, with
the present available computer this regime is not fully accessible to us.
From the figure it is quite clear for smaller lattices ($80\times80$), that
one has to go below the frequency range $\omega \simeq 10^{-6}$, to get the
results for hysteresis which will be comparable with that obtained from
classical nucleation theory. For the large lattice sizes, this value of
the frequency will be much smaller. For example, if $L = 1200$, one can 
see (from extrapolation) that the crossover might be at $ \omega 
\simeq 10^{-7}$.
For shorter times (high frequency), the simulation results and 
the theoretical prediction (from CNT)
for the coercive field (or loop area) with respect to frequency
will disagree with each other. Due to this reason, the recent simulation
results \cite{ma} show a power law variation of the coercive field
with respect to frequency which is not in agreement with the theoretical
prediction \cite{dhar} obtained from CNT.

The simulations (for static field) are also done in $d=3$ and 4. 
The results are depicted 
in Fig. 4 and Fig. 5 respectively. 
Here also we observed the results are consistent with that of CNT.
In table I, all the results are summarised.
The computational time for the largest lattices in all three dimensions are
also given there. The surface tensions, calculated from the simulation
($\sigma_{Sim}$) are compared with the previous estimates ($\sigma_{Prev}$).


\begin{center} {\bf Table I} \end{center}
\begin{center}
\begin{tabular}{|c|c|c|c|c|c|c|c|c|}
\hline
$d$ & $L_{max}$ & $\tau_{max}$ & $N_p$& Time& $J/K_BT$& Slope& $\sigma_{Sim}$&
$\sigma_{Prev}$ \\
& & & &(CPU)& &(in coalsc. & &\\
& & & & & & regime) & &\\
\hline
 2 & 2048& 9160000 &512 &5293.3 Sec.&0.625&0.27&1.378&$\sigma_{Hor} \simeq 
$1.06 \\
\hline
 3 & 256& 144250 &32 &5864.8 Sec.&0.375&0.70&1.454&$\simeq 1.3$ \\
 &&&&&& && Ref.\cite{burkner}\\
\hline
 4 & 48& 55750 &16 &2602.9 Sec.&0.220&1.10&1.363&-- \\
 &&&&&& &&\\
\hline
\end{tabular}
\end{center}

\noindent In the above table $\sigma_{Hor}$, the surface tension (in units of
$J$) for horizontal direction, has been calculated from Ref. \cite{wort} and 
agrees nearly with the
surface tension for 45$^o$ direction with respect to the 
horizontal line. See ref \cite{rikvold}a
 for a discussion of the discrepancy between
$\sigma_{Sim}$ and $\sigma_{Prev}$

\bigskip
\leftline {\bf IV. Summary}

We have studied nucleation in the two-, three- and four- 
dimensional Ising system by
Monte Carlo simulation with heat-bath dynamics. The logarithm of
the nucleation time is found to be proportional to the 
($-(d-1)$) power of the magnetic field.
The size-dependent crossover from coalescence to nucleation
is observed clearly
in all dimensions. The surface tensions have been estimated from
the proportionality constants (related to the surface tension) and are
compared with the previous bulk estimates.
The results are roughly consistent
with the prediction of classical nucleation theory.

In two dimensions, the dynamical responses of the system are studied in 
a sinusoidally oscillating magnetic field. The reversal time
has been studied as a function of the coercive field. For low enough frequency
the logarithm of the reversal time is found to be proportional to the
inverse of the coercive field. These results are compared with the nucleation
results for static field; in the nucleation regime, the reversal time
and the nuclation time become identical for low enough frequency.

However, for intermediate frequencies,
the Monte Carlo results 
\cite{ac,ma}
for the frequency variation of the dynamic
coercivity and the hysteresis loop area do not agree with the theoretical
predictions 
\cite{dhar} obtained from classical nucleation theory.

In a recent study \cite{mc} the spin reversal transition was found in 
the two dimensional kinetic Ising model in a short-duration pulsed
magnetic field. The phase boundary was drawn in the plane formed by the
strength and duration (of activity) of the field. This is nothing but
the variation of nucleation time as a function of field-strength. However,
in that study, the results are mostly confined to strong fields and 
the coalescence regime. 
The asymptotic ($\Delta \to \infty$) functional
form of this phase boundary can also be predicted from the classical nuclation
theory and it will be $\ln(\Delta t) \sim 1/h_p$ for very large $\Delta t$ and
small $h_p$.

Recently \cite{uli} a similar crossover from nucleation regime 
to coalescence regime has
been observed by Monte Carlo simulation in three-dimensional 
anisotropic (large) Heisenberg model 
by tuning the temperature.

\bigskip
\noindent {\bf Acknowledgments:} Sonderforschungsbereich 
341 is gratefully acknowledged by one of us (MA) 
for financial support. MA would like to thank Bikas
Chakrabarti, Deepak Dhar and Per Rikvold for important
discussions. Authors would like to acknowledge the computational facilities
provided by HLRZ, J\"ulich, Germany.

\bigskip

\begin{center} {\bf Figure captions} \end{center}

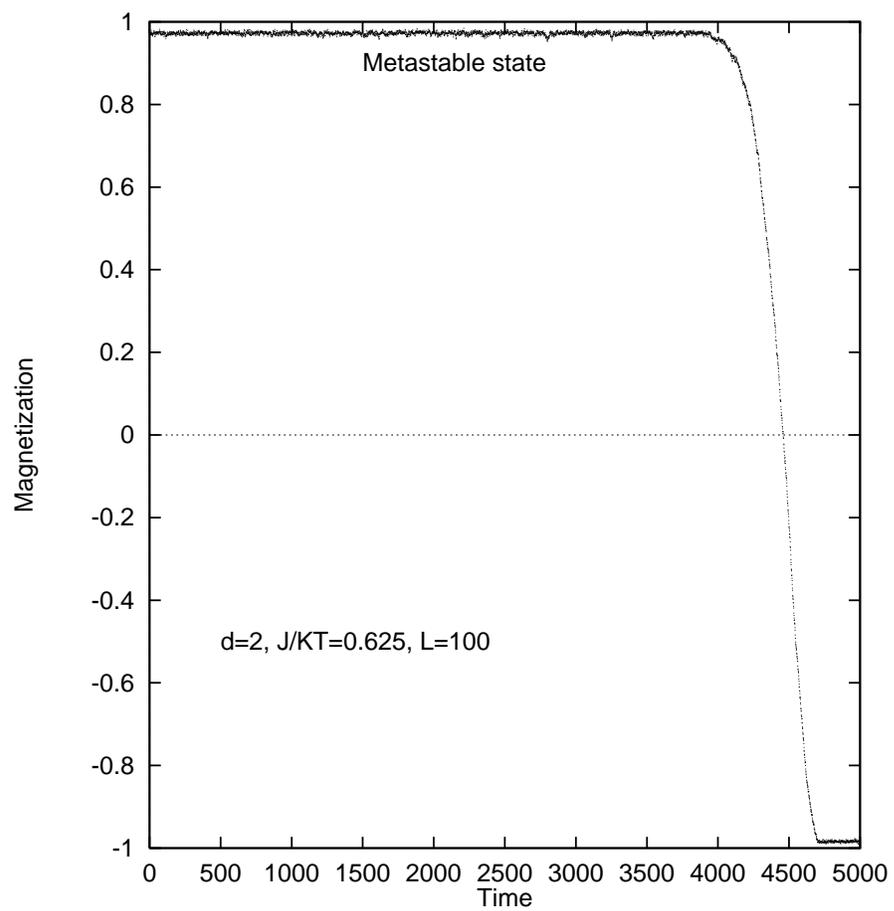
\noindent {\bf Fig. 1.} A typical decay of metastable state.

\bigskip

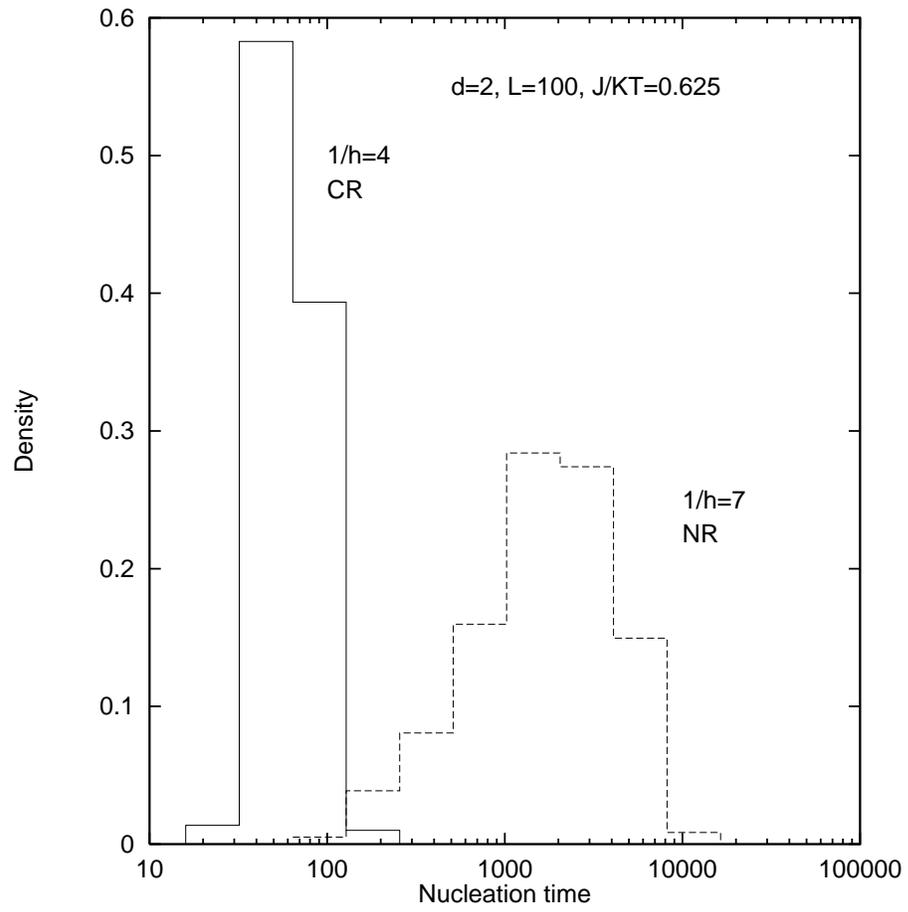
\noindent {\bf Fig. 2.} Distribution of nucleation times in coalescence
regime (CR, $1/h=4$) and in nucleation regime (NR, $1/h=7$). 
The results are obtained for $L = 100$ in $d = 2$.

\bigskip

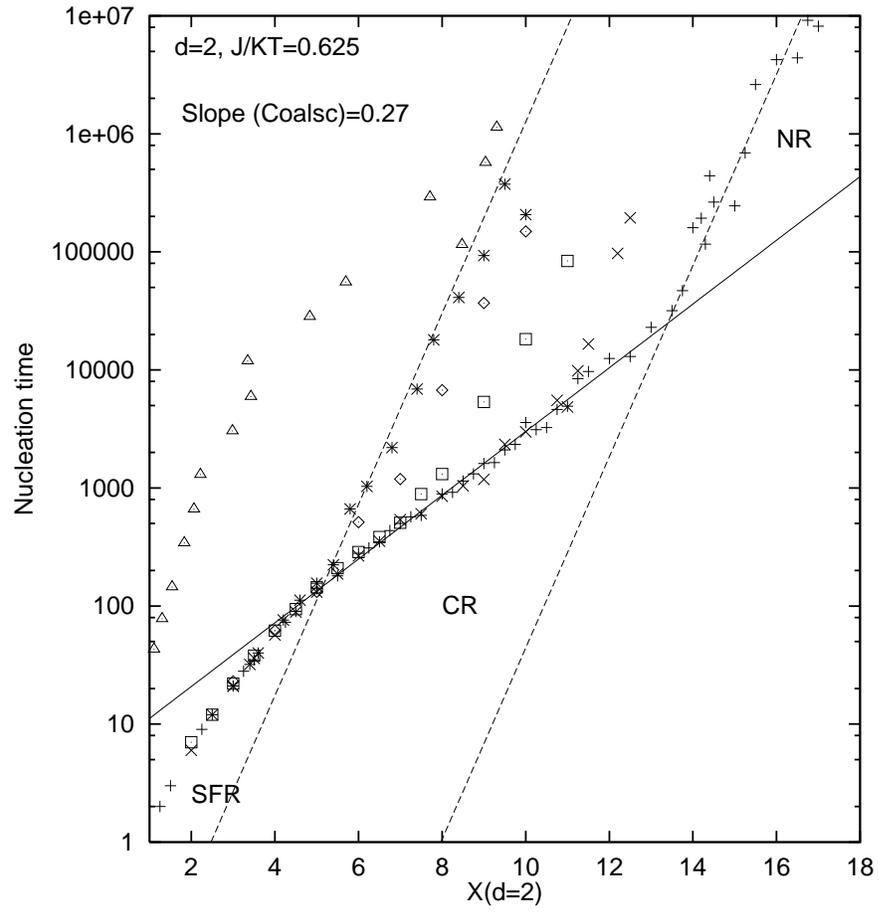
\noindent {\bf Fig. 3.} Nucleation time ($\tau$) 
(in logscale) plotted against $X(d=2)= 1/h$ in
$d = 2$. Different symbols correspond to different system sizes. $L = 80$
 (star), $L=100$ (diamond), $L=400$ (box), $L=1200$ (cross) and $L=2048$ 
(plus). The solid line is the linear best fit in the coalescence regime (CR).
The dashed line has a slope three times higher than that of 
the solid line. The
reversal time (triangles for $L$ = 80; in logscale)
is also plotted against the inverse of the coercive field in the
same plot. The nucleation regime is also shown here
by another dashed straight line. In the low field regime, the reversal time and
the nucleation time follow the same behaviour and are expected to agree.
The intercepts of the straight lines depend on $L$.

\bigskip

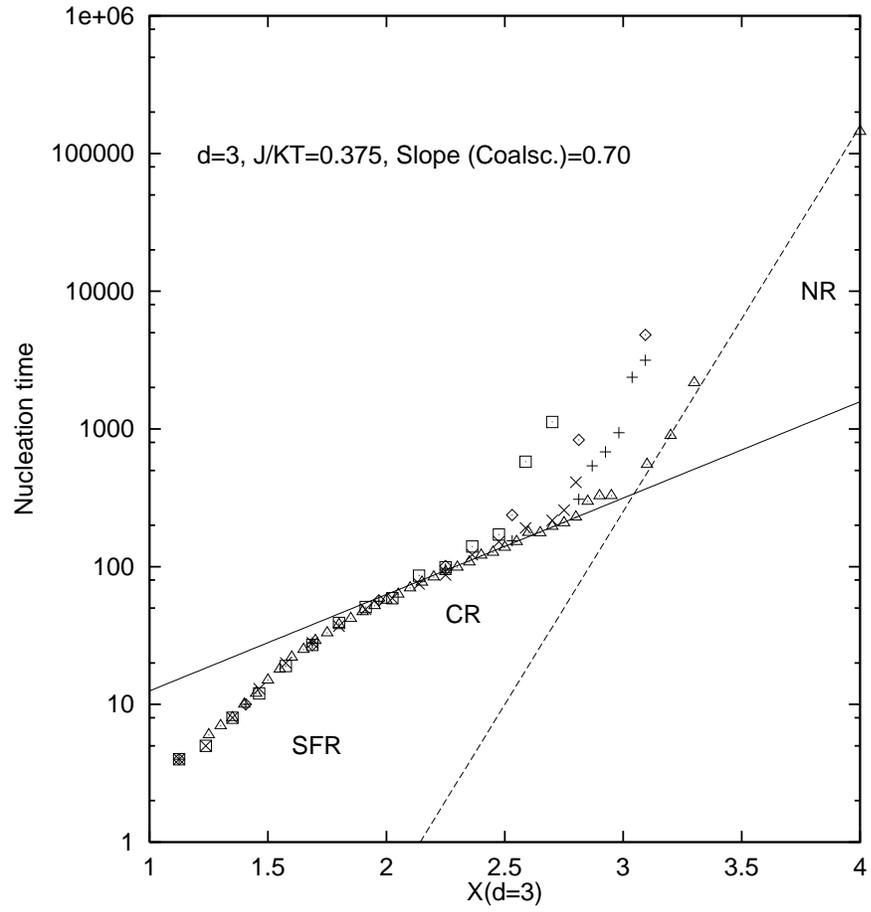
\noindent {\bf Fig. 4.} Nucleation time ($\tau$) 
(in logscale) plotted against $X(d=3) = 1/h^2$ in
$d = 3$. Different symbols correspond to different system sizes. $L = 60$
 (box), $L=64$ (diamond), $L=90$ (cross), $L=128$ (plus) and $L=256$ 
(triangle). The solid line is the linear 
best fit in the coalescence regime (CR).
The dashed line has a slope four times higher than that of the solid line. 

\bigskip

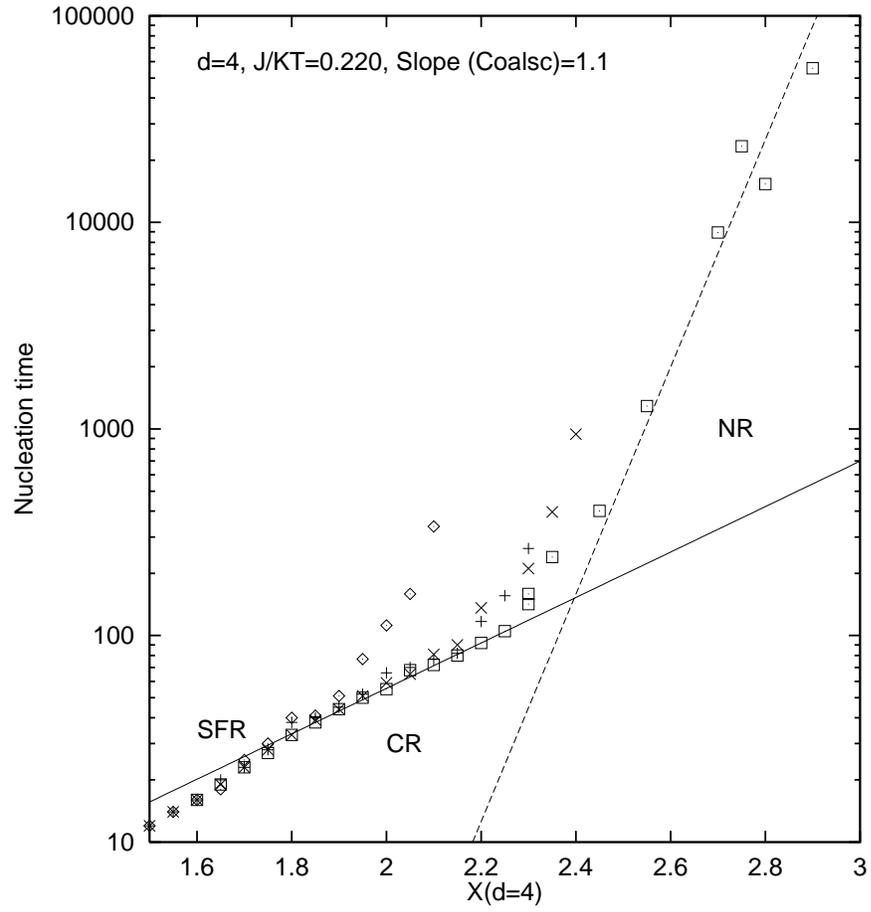
\noindent {\bf Fig. 5.} Nucleation time ($\tau$) 
(in logscale) plotted against $X(d=4)= 1/h^3$ in
$d = 4$. Different symbols correspond to different system sizes. $L = 17$
(diamond), $L=31$ (plus), $L=32$ (cross) and $L=48$ (box). 
The solid line is the linear best fit in the coalescence regime (CR).
The dashed line has a slope five times higher than that of the solid line. 
\end{document}